# Ferrotronics for the creation of band gaps in Graphene

Q. Wan[1], Z. Xiao, A. Kursumovic[2], J. L. MacManus-Driscoll[2] & C. Durkan[*, 1]

[1]: Nanoscience & Department of Engineering

University of Cambridge

11 JJ Thomson Avenue

Cambridge, CB3 0FF

UK

[2]: Department of Materials Science & Metallurgy,

University of Cambridge

27 Charles Babbage Road

Cambridge, CB3 0FS

UK

*: corresponding author, email cd229@cam.ac.uk

*We experimentally demonstrate a simple bilayer graphene/ ferrolectric device, termed Ferrotronic (electronic effect from ferroelectric) device in which the band-structure of single-layer graphene is modified. The device architecture consists of graphene deposited on a ferroelectric substrate which encodes a periodic surface potential achieved through domain engineering. This structure takes advantage of the nature of conduction through graphene to modulate the Fermi velocity of the charge carriers by the variations in surface potential, leading to the emergence of energy mini-bands and a band gap at the superlattice Brillouin zone boundary. Our work represents a simple route to building circuits whose functionality is controlled by the underlying substrate.*

## Introduction

The energy spectrum of graphene is linear and gapless in the vicinity of the Dirac (K & K') points. The charge carriers have spin ½ and are described by the massless Dirac equation: $E = \pm \hbar v_F \vec{k}$ with a Fermi velocity, $v_F \sim 10^6$ m/s and mean free path often far in excess of that in metals[1]. Many attempts have been made to exploit this high velocity in a range of electronic and optoelectronic applications, including interconnects, transistors, sensors and saturable absorbers[2–5]. However, a fundamental issue that limits the use of graphene in practical devices is the fact that it is gapless. This leads to an impractically low on/off ratio for graphene field-effect transistors (GFETs), typically no more than 10 - 20, which is substantially smaller than that which is found in conventional semiconductor FETs, of the order of $10^7$ or higher. Therefore, in an attempt to address this, there is a growing interest in exploring ways of introducing a bandgap. For bilayer graphene, this can be achieved by breaking the system symmetry, for instance by application of an electric field perpendicular to the layers[6], but for monolayer graphene, no such phenomenon exists. One option is via the quantum size effect – graphene nanoribbons (GNRs) have a width-dependent bandgap in excess of 0.1 eV for ribbon widths below 40 nm[7]. It has been shown theoretically that graphene's band structure can be modified in the presence of a periodic potential- known as electrostatic superlattices[8–11], reminiscent of the nearly-free electron (e.g. Kronig-Penney [12]) model. One of the effects of periodic potentials on graphene's band structure is to introduce an energy gap at the superlattice Brillouin Zone (SBZ). In this work, we report on a device configuration where we apply a periodic potential with multiple harmonics to a graphene monolayer using ferroelectric domains in an underlying film to create a series of mini bandgaps at the relevant SBZ points. Our device structure has major advantages over previous work on bandgap engineering of graphene, which relied on ultrahigh resolution lithography and patterning of GNRs[7], which suffers from far less control due to issues around variations in feature size leading to significant fluctuations in band gap. Our approach has the benefit that in principle, the band gap is set via the substrate surface potential, which in principle can be achieved using other means than ferroelectric domains, and this can be controlled at will.

## Methods

The periodic electrostatic potentials described above are created by means of a poled ferroelectric substrate. Figure 1(a) shows the overall device layout of our PZT-graphene integrated device. The ferroelectric material PZT (lead zirconate titanate) was PVD-grown on $SrTiO_3$ substrates to induce a (100) orientation, in order to ensure the resulting electric field from the polarisation is normal to the surface. PZT was chosen as it has the highest surface polarisation and electromechanical coupling of any ferroelectric material and is widely used and accepted by industry. The 1D periodic potential was patterned using voltage-controlled domain poling[13] by means of an AFM (Atomic Force Microscope). The presence of the patterned ferroelectric domains was then verified using PFM[14] (Piezoresponse Force Microscopy), as shown in Figure 1(b), where the period ($L$) in this particular case is 80 nm. The smallest 1D periodic potential period that was successfully patterned and subsequently imaged in this way in these films is 50 nm. The ferroelectric domains are vertically-oriented, alternating between polarisation direction pointing up and down, associated with a surface potential which periodically alternates between positive and negative, respectively. This can be thought of as a 1D square wave potential distribution - an electrostatic superlattice. KPFM[15] (Kelvin Probe Force Microscopy) was used to determine the difference in surface potentials of the polarised-up and polarised-down regions, i.e. the height of the square barriers ($V_0$), which were found to be in the range 50 meV – 300 meV. This periodic variation in surface potential in turn gives rise to periodic variations in the Fermi level of graphene placed on top of the PZT surface. A device is then completed with the addition of a pair of electrodes – one at either end of the superlattice. The conductive substrate acts as a back-gate, so the overall device is essentially a hybrid graphene/ferroelectric FET. Figure 1(c) shows an SEM image of the active region of a device, where the distance between the electrodes is 310 nm. The processing steps that we employ for device fabrication tend to lead to an electronic mean free path in the range 100-220 nm at room temperature[16]. At low temperatures, this is expected to exceed the size of our devices, resulting in coherent electron transport throughout.

## Findings & Discussion

Figure 2a shows the transfer characteristics (Drain current *vs.* gate voltage) of such a device, where the conductance near the Dirac point deviates from the usual smooth, near-parabolic dependence on voltage seen in a normal graphene FET. There is a flat region, in this case 340 meV wide. The presence of the superlattice leads to the formation of minibands and energy gaps at minizone boundaries, which collectively lead to a net reduction in conductance in this region. Figure 2b demonstrates the temperature dependence of the electrostatic graphene superlattice effect. Gate sweep measurements were carried out at 10K, 20K and 30K. At 10K, a flat region featuring periodic variations in conductance is seen near the Dirac point. At elevated temperatures starting from around 30K onwards, these variations become less apparent and the characteristic becomes rather similar to that of a normal graphene FET[17, 18].

A detailed knowledge of the form of the potential allows us to predict where the SBZ points will lie. The 1D periodic potential in this work is a square wave of amplitude $U_{1D}$ and period $L$, and can be written in the form of a Fourier series as

$$f(x) = U_{1D} \sum_{n=1,3,5....}^{\infty} \frac{2}{n\pi} \sin\left(\frac{2n\pi}{L}x\right). \qquad (1)$$

This comprises only odd harmonics, and the key point is that they will lead to a series of minizone boundaries at $k_x = \pm\pi/L, \pm3\pi/L, \pm5\pi/L$, etc. The size of the bandgap at each minizone boundary has been calculated in terms of the superlattice characteristics as[8]

$$\Delta E = \frac{2}{\pi} U_{1D} \sin\theta_{\vec{k},\hat{x}}. \qquad (2)$$

where $\theta_{\vec{k},\hat{x}}$ is the angle between the wavevector $\vec{k}$ $(k_x, k_y)$ and direction $\hat{x}$ - the direction of the periodic potential.

For purely normal incidence of electrons on the superlattice (in which case $\theta_{\vec{k},\hat{x}} = 0$), Klein tunnelling leads to perfect transmission of electrons owing to the suppression of back-scattering processes for massless Dirac fermions. In this case, no bandgap will be observed by those electrons. However, in reality the current will flow at a range of angles between the drain and source electrodes such that the incidence on the barriers is a mixture of normal and oblique incident angles. Therefore, there is a possibility that a proportion of electrons will be confined in the potential wells between the barriers, and the spectrum of those bound states give rise to multiple minibands. The graphene superlattice effect can be probed via measurements of the current ($I_d$) through the PZT-graphene integrated device or the corresponding conductance ($G_d$) as a function of the relative gate voltage ($V_g$ - $V_{CNP}$), at a fixed drain-source voltage ($V_{DS}$). For the measurements reported here, $V_{DS}$ was held at 1 mV.

Using the known intrinsic band structure of graphene, we can predict the energies (relative to the Dirac point) at which band gaps would be expected, given that they occur at the above values of $k_x$, assuming a fixed value for $k_y$, as illustrated in Figure 3a. The values of these bandgaps: $\Delta E_1$, $\Delta E_2$, $\Delta E_3$ etc are calculated from equation (2). The segment between $E_n$ and $E_n + \Delta E_n$ represents the energy gap between two adjacent minibands created by the superlattice. Since the data collected from the experiments are gate sweep measurements, the energy $E$ and energy gap $\Delta E$ are then converted to gate voltage using the dispersion relation (from Figure 3a), as illustrated in Figure 3b. It should be noted that all calculations involved here are carried out with the gate voltage adjusted by the charge neutrality point (CNP) value. The segment of the gate sweep curve between $\frac{E_n}{e}$ and $\frac{E_n}{e} + \frac{\Delta E_n}{e}$ corresponds to the superlattice - induced energy gap between two neighbouring minibands in Figure 3a. By mapping the created minibands from the dispersion relation onto the gate sweep curve, one is able to construct the theoretically predicted effect from a 1D graphene superlattice on a $G_d$ versus ($V_g$ - $V_{CNP}$) map.

The experimentally measured and theoretically predicted locations of the minizone boundaries (MB) as well as the corresponding energy gaps are shown in Figure 4 as shaded boxes, the left side of which coincides with the minizone boundary and the span of which refers to the energy gap between adjacent minibands. The conductance will drop within energy gaps and increase again outside of them. Therefore, we would expect the transfer characteristics of a device to display variations in conductance ($G_d$), as minibands are opened and closed.

In the particular case shown in Figure 4, the superlattice has parameters $U_{1D} = 0.15$eV, $L = 60$nm. The best fit between theory and experiment occurs when $k_y = 0.6$Å$^{-1}$, corresponding to $\theta_{\vec{k},\hat{x}} = 89.5°$. Under the above superlattice conditions, the first energy gap is large enough to extend into the second and third minibands, such that the second and third minibands start before the energy gap between the first and second minibands finishes. The conductance drop locations No. 1, 2, 3, 5, 6 in Figure 4 can be matched with that predicted by theory, while location No.4 is within a predicted miniband between two shaded boxes. In all cases, we obtain the best agreement between theory and experiment for the first 5-6 minibands (blue dashed lines), beyond which they start to deviate.  This is to be expected as the superlattice is finite in extent, comprising approximately 5 periods, so the above Fourier series approach is only going to be accurate for the first few harmonics.  A point to note is that the variations in conductance are especially small, and certainly smaller than would be the case if the charge carriers in graphene were to follow Schrödinger's equation, i.e. if they were not effectively massless Dirac Fermions.  A simple model of such a structure leads to a transmission probability which varies over a significantly larger range, as there is no Klein tunnelling.  The fact that the current variation we observe is so low is further indirect evidence for Klein tunnelling, in line with measurements by others[19].

## Conclusion

We have experimentally demonstrated that the influence of an electrostatic superlattice on graphene is to generate minibands and mini-bandgaps, in line with theoretical predictions. Periodically-poled ferroelectric substrates were used to create a substrate with alternating polarities of surface potential in the range 50 – 300 meV, which was observed to modify the band structure of monolayer graphene placed on top.  This combination of graphene and ferroelectric substrate, which we term "Graphene Ferrotronics" was demonstrated to display a series of mini-bandgaps whose position and size depended on the period and amplitude of the periodic electrostatic potential.  This paves the way towards advanced device fabrication using fewer materials than conventional CMOS processes, whereby specific properties can be achieved through substrate patterning.

## Competing interests

The authors declare that there are no competing interests.

## Author contributions

The concept behind and design of the experiment was carried out by CD.  The experiments were carried out by QW.  The processes for patterning and cleaning graphene were developed by ZX. The Ferroelectric samples were grown by AK and JMD.  The paper was written by QW, CD and JMD with contributions from ZX and AK.

## Data availability

The data that support the findings of this study are available from the authors on reasonable request

Figure Captions

**Figure 1.** Hybrid Graphene/ferroelectric G-FET Device configuration. (a) Device schematic showing graphene strip on periodically-poled ferroelectric substrate, and with source, drain and gate connections; (b) PFM image showing ferroelectric domains between source and drain contacts (which are at the top and bottom of the imaged strip, respectively); (c) SEM image of device showing electrical contacts on top of graphene strip.

**Figure 2.** Conductance measurements demonstrating existence of a bandgap. (a) Conductance Vs relative gate voltage of a graphene/ferroelectric device, where a flattened region of width ~ 340 meV can be seen around the Dirac point; (b) Conductance Vs relative gate voltage as a function of Temperature, where a series of mini band gaps are apparent up to 20K, but not above that. These band gaps overlap to produce the flat region seen in Figure (a).

**Figure 3.** Location of superlattice Brilluoin zones (SBZ). (a) Band structure of graphene, whereby the locations of the SBZ are indicated on the x (k) axis showing wave-vector, on the basis of the harmonics associated with the square-wave potential. On this basis, we can determine the values of Energy (related to Gate Voltage) where band gaps should appear; (b) mapping the locations of these band gaps onto a conductance measurement. The expected width of the band gaps decreases with increasing harmonic as expected for a square wave.

**Figure 4.** Mapping of expected locations of mini band gaps associated with SBZ, from Figure 3. Zones 1-6 correspond to SBZ band gaps 1-6, respectively. The proximity of these bandgaps to each other and their partial overlap is what gives rise to the flattened region around the Dirac point in the conductance-relative gate voltage characteristics.

# Figure 1

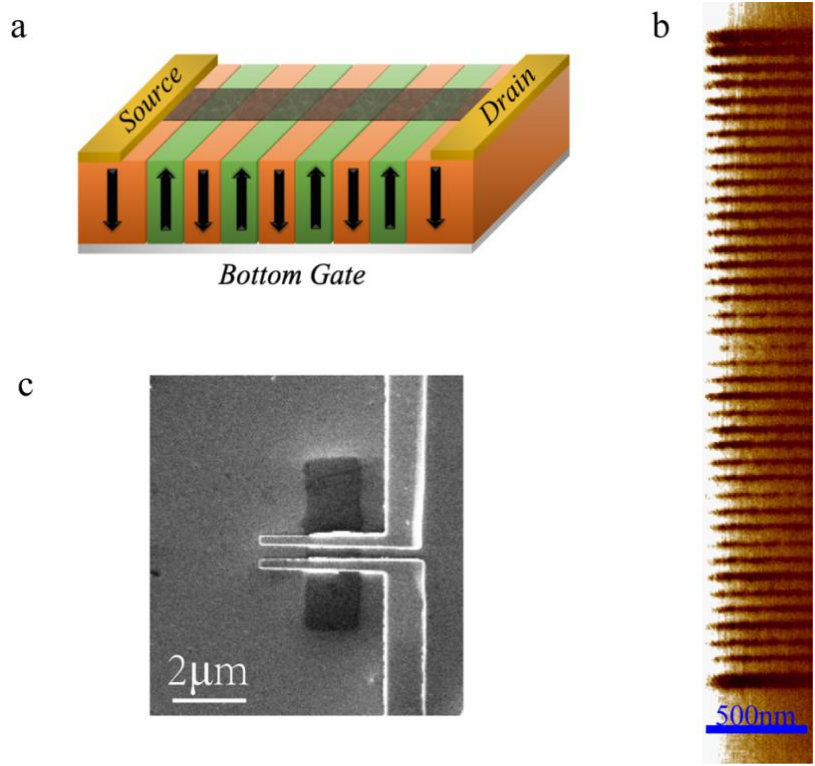

**Figure 2**

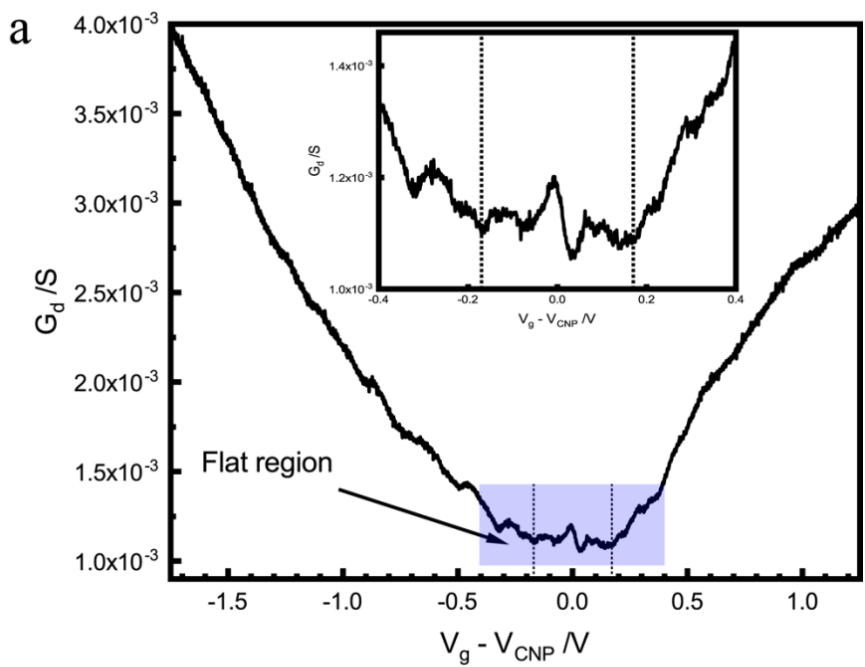

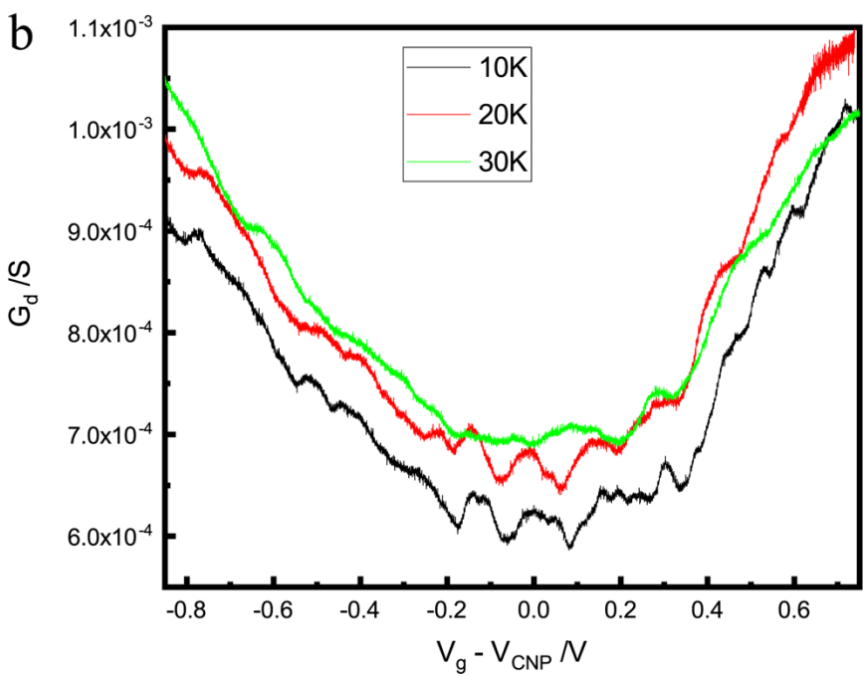

**Figure 3**

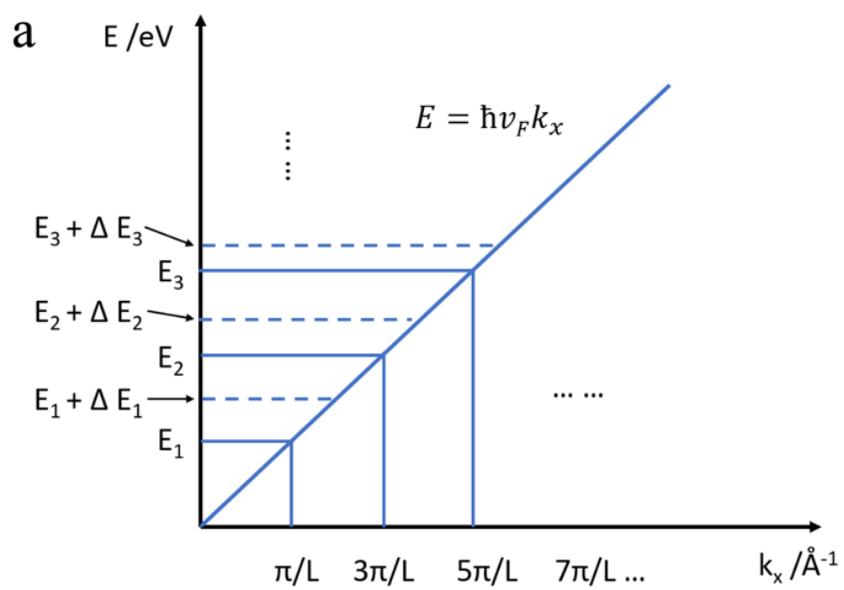

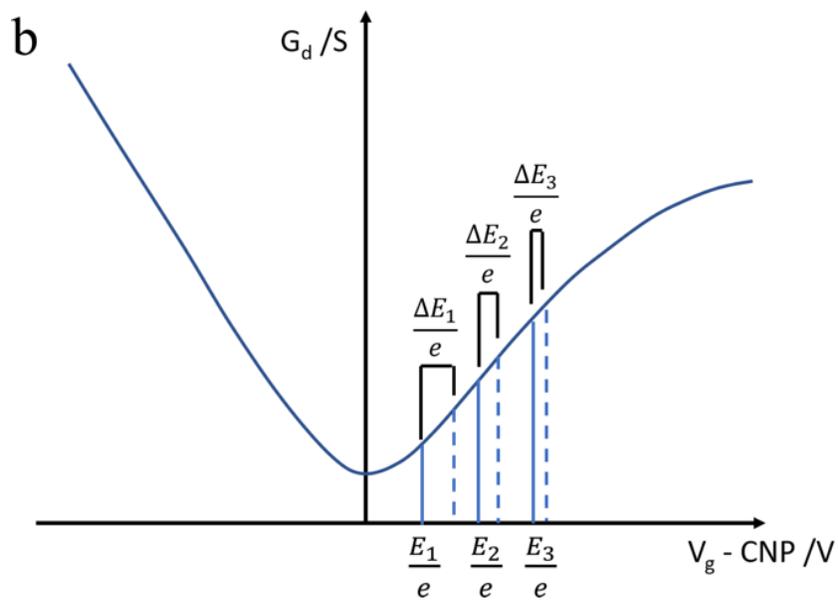

**Figure 4**

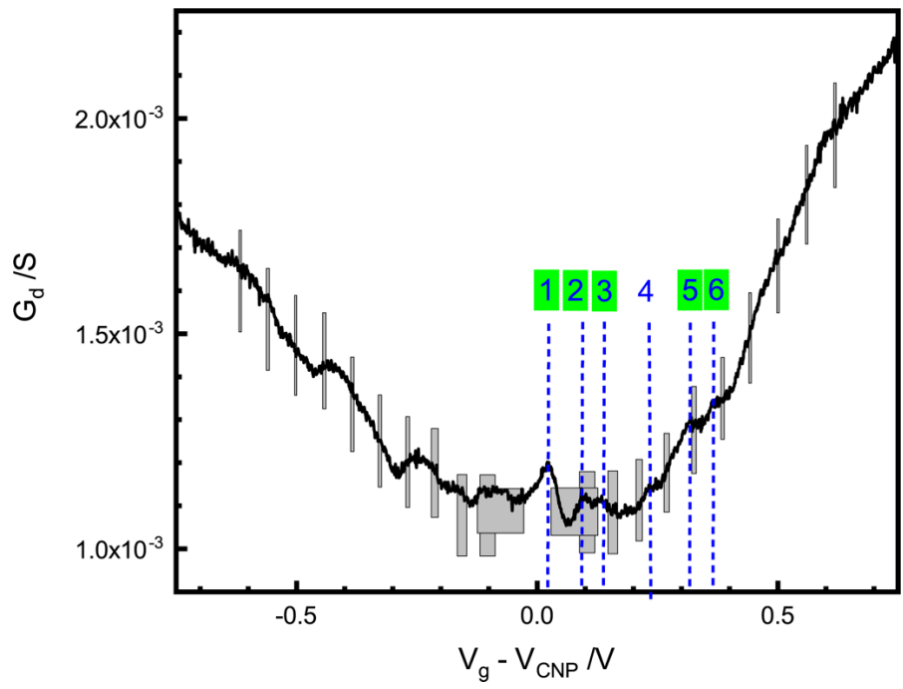